\documentclass[12pt]{JHEP3}

\usepackage{amsmath}
\usepackage{amsfonts}
\usepackage{amssymb}
\usepackage{graphicx}
\usepackage{epstopdf}
\usepackage{epsfig}

\newcommand{\tr}{\mathrm{tr}}

\newcommand{\comma}{\; ,}

\title{Quenched mesonic spectrum at large $N$}

\author{Luigi Del Debbio \\
  SUPA, School of Physics and Astronomy, University of Edinburgh\\
  Edinburgh EH9 3JZ, Scotland\footnote{Permanent address.},\\
  {\rm and} Isaac Newton Institute for Mathematical Sciences\\
  20 Clarkson Road, Cambridge CB3 0EH, UK\\
  E-mail: \email{luigi.del.debbio@ed.ac.uk} 
} 
\author{Biagio Lucini \\
  Physics Department, Swansea University\\
  Singleton Park, Swansea SA2 8PP, UK$^{\ast}$,\\
  {\rm and} Isaac Newton Institute for Mathematical Sciences\\
  20 Clarkson Road, Cambridge CB3 0EH, UK\\
  E-mail: \email{b.lucini@swansea.ac.uk} 
}
\author{Agostino Patella \\
  Scuola Normale Superiore, Piazza dei Cavalieri 27, 56126 Pisa, Italy\\ 
  and INFN Pisa, Largo B. Pontecorvo~3~Ed.~C, 56127 Pisa, Italy\\
  E-mail: \email{agostino.patella@sns.it} 
} 
\author{Claudio Pica \\
  Physics Department, Brookhaven National Lab\\
  Upton, NY 11973-5000, USA\\
  E-mail: \email{pica@bnl.gov} 
}

\abstract{We compute the masses of the $\pi$ and of the $\rho$ mesons
in the quenched approximation on a lattice with fixed lattice spacing
$a \simeq 0.145 \ \mathrm{fm}$ for SU($N$) gauge theory with $N = 2,3,4,6$.
We find that a simple linear expression in $1/N^2$ correctly captures the
features of the lowest-lying meson states at those values of $N$. This enables
us to extrapolate to $N = \infty$ the behaviour of $m_{\pi}$ as a function of
the quark mass and of $m_{\rho}$ as a function of $m_{\pi}$. Our results for
the latter agree  within 5\% with recent predictions obtained in the
AdS/CFT framework.}

\preprint{BNL-NT-07/56\\IFUP-TH/2007-36\\NI07089}


\begin{document}

\section{Introduction}
\label{sec:intro}
The theory of strong interactions, Quantum Chromodynamics (QCD), is a
SU(3) gauge theory with $n_f$ flavors of fermionic matter fields in
the fundamental representation of the color group. For sufficiently
small $n_f$, QCD displays many interesting non--perturbative
phenomena, which are not captured by the conventional expansion in
powers of the coupling constant. However, if we consider a SU($N$)
gauge theory, with a generic number of colors $N$, in the limit where
$N$ becomes large, the perturbative expansion can be reorganized in
powers of $1/N$, and the contribution of each diagram can be directly
related to its topology~\cite{'tHooft:1973jz,Veneziano:1974ag}. The
leading contribution in this expansion is given by planar diagrams,
and a simple power counting argument suggests that corrections are
${\cal O}(1/N^2)$ in a pure gauge theory, while the fermionic
determinant yields corrections ${\cal O}(n_f/N)$.

The large--$N$ expansion is a powerful tool to explore the strongly 
interacting regime of gauge theories, and recent developments in string
theory have provided beautiful insights in our understanding of the planar
limit through the gauge-gravity correspondence~\cite{Maldacena:1997re}
(see~\cite{Nastase:2007kj} for an introductory review of recent developments).
The lattice formulation of gauge theories allows one to study the
non-perturbative dynamics from first
principles by numerical simulations, and can therefore be used to
investigate how the $N=\infty$ limit is approached. A number of studies in
recent years~\cite{Lucini:2001ej,Lucini:2003zr,Lucini:2004yh,DelDebbio:2001sj,DelDebbio:2002xa,DelDebbio:2004rw,DelDebbio:2006df}
have analyzed in detail several features of pure gauge theories for
$N\geq 2$, including the spectrum of glueballs, the $k$-string
tension, and topology, both at zero and finite temperature. A very
precocious scaling has been observed for all observables that have
been considered so far, with $1/N^2$ corrections being able
to accommodate the values of the observables already for $N=3$ and in most
of the cases also for $N = 2$.

The convergence to the large--$N$ limit for theories with fermions
could also be addressed by dynamical simulations. The contributions of
the fermionic determinant should increase the size of the corrections,
as pointed out above. An intermediate step at a lesser computational
cost is the study of properties of mesons and baryons in theories with
quenched fermions. Note that since the fermionic determinant is
suppressed in the large--$N$ limit, simulations in the quenched
approximation should converge to the same limit as in the theory with
dynamical fermions, but with corrections ${\cal O}(1/N^2)$.

This paper focuses on the low--lying states of the mesonic spectrum
for SU($N$) theories in the quenched approximation and $N=2,3,4,6$.
By generalizing the lattice Dirac operator to handle spinors of
arbitrary dimension in color space, we compute two-point functions for
Wilson fermions at one value of the lattice spacing and several values
of the bare quark mass. The mass dependence of the spectrum is
studied, and extrapolated to the large--$N$ limit. Our results are
consistent with a $1/N^2$ scaling, and the results for $N=\infty$
can be used as an input for analytical approaches that study the meson
spectrum of strongly--interacting gauge theories. Some aspects of the
meson spectrum at large $N$ (in particular, the dependence of the
mass of the pion from the quark mass) have also been investigated
in~\cite{Narayanan:2005gh}. 

We stress that this calculation is meant to be exploratory, trying to
favor a first overall physical picture over more formal and technical
points. A more detailed calculation is currently in progress and will
be reported elsewhere.

The paper is organized as follows. Sect.~\ref{sec:lat} recalls the
basic framework that is used for extracting the mesonic spectrum from
field correlators in quenched lattice gauge theories and summarizes
the choice of bare parameters for each value of $N$. The
numerical results and their analysis are presented in
Sect.~\ref{sec:num}. Finally, we conclude by discussing the large--$N$
extrapolation in Sect.~\ref{sec:extr} and its relevance for AdS/QCD studies
(see~\cite{Erdmenger:2007cm} for a review) in Sect.~\ref{sec:dis}.

As this work was being completed we noticed that similar problems have
been investigated in Ref.~\cite{Bali:2007kt}. The preliminary results
presented there are obtained on slightly smaller lattices (in physical
units) with a finer lattice spacing. The two sets of data are
complementary and in qualitative agreement. Future calculations will
hopefully achieve precise continuum results for the large--$N$ limit of
the mesonic spectrum.
\section{Lattice formulation}
\label{sec:lat}
A Monte Carlo ensemble of gauge fields is generated using the Wilson
formulation of pure SU($N$) gauge theory on the lattice, defined by the
plaquette action
\begin{equation}
  \label{eq:wilson}
  S=-\frac{\beta}{2 N}\sum_{x,\mu>\nu} \mathrm{Tr}\left[U(x,\mu) U(x+\mu,\nu)
  U^\dagger(x+\nu,\mu) U^\dagger(x,\nu) + \mathrm{h.c.}\right] \ ,
\end{equation}
where $U(x,\mu)\in \mathrm{SU}(N)$ are the link variables. The link
variables are updated using a Cabibbo--Marinari
algorithm~\cite{Cabibbo:1982zn}, where each SU(2) subgroup of SU($N$) is
updated in turn. We have alternated microcanonical and heat--bath steps in a
ratio 4:1. We call {\em sweep} the sequence of four microcanonical and one
heat--bath update.

The action of the massive Dirac operator on a generic
spinor field $\phi(x)$ is:
\begin{eqnarray}
D_m \phi(x) &=& (D+m) \phi(x) \nonumber \\
&=& - \frac12 \left\{ \sum_{\mu} \left[ \left(1-\gamma_\mu\right)
U(x,\mu) \phi(x+\mu) + \left(1+\gamma_\mu\right) U(x-\mu,\mu)^\dagger \phi(x-\mu)\right] - \right. 
\nonumber \\
& & \left. -(8+2m) \phi(x) \right\}.
\end{eqnarray}
The bare mass is related to the hopping parameter used in the actual
simulations by
\begin{equation}
  \label{eq:hopk}
  1/(2\kappa)=4+m \ .
\end{equation}
The complete set of bare parameters used in our simulations is
summarized in Tab.~\ref{tab:par}. For this preliminary study we have
run simulations for values of $N$ ranging from 2 to 6 at one value of
the lattice spacing only. The values of $\beta$ have been chosen in
such a way that the lattice spacing is constant across the various
$N$.  More in detail, at each $N$ we chose the critical value of
$\beta$ for the deconfinement phase transition at $N_t =
5$~\cite{Lucini:2004yh}. To set the scale, another physical quantity
(like e.g. the string tension) can be used. Since different
quantities have different large--$N$ corrections, a different choice for
the scale will affect the size of the $1/N^2$ corrections, but not the
$N = \infty$ value. Using the value $T_c = 270 \ \mathrm{MeV}$,
for the lattice spacing $a$ we get $a \simeq 0.145 \ \mathrm{fm}$. For
our simulations, we used a $N_s^3 \times N_t$ lattice with the spatial
size $N_s = 16$ (which corresponds to about $2.3 \ \mathrm{fm}$) and
the temporal size $N_t = 32$ (about $4.6 \ \mathrm{fm}$ in physical
units). For the quark masses used in this work, our calculation should
be free from noticeable finite size effects. For the fermion fields we
used periodic boundary conditions in the spatial directions and
antiperiodic boundary conditions in the temporal direction. For the
gauge field we used periodic boundary conditions in all directions.

\TABLE{
\begin{tabular}{|c|r|c|}
\hline
N & $\beta$ & $\kappa$ \\
\hline
2 & 2.3715 & 0.156, 0.155, 0.154, 0.153, 0.152\\ 
3 & 5.8000 & 0.161, 0.160, 0.159, 0.1575, 0.156\\ 
4 & 10.6370 & 0.161, 0.160, 0.159, 0.1575, 0.156\\ 
6 & 24.5140 & 0.161, 0.160, 0.159, 0.1575, 0.156\\ 
\hline
\end{tabular}
\caption{Bare parameters used in our simulations.}
\label{tab:par}
}

We have performed a chiral extrapolation using data from meson
correlators at five values of $\kappa$ for each $N$. The choice of
values for $\kappa$ relies on previous experience with SU(3)
simulations to yield pseudoscalar meson masses $\ge 450 \
\mathrm{MeV}$.
The same values of $\kappa$ have been used for all
values of $N \ge 3$, while for SU(2) a different choice turned out to
be necessary, since all $\kappa$'s but the lowest one were higher than
$\kappa_c$.  Because of the different additive renormalization, these
values of $\kappa$ yield different values for the bare PCAC mass as
$N$ is varied.

The mesonic spectrum is extracted from the zero-momentum two-point
correlators of quark bilinears with the quantum numbers required to
interpolate between a meson state and the vacuum. Let $\Gamma_1$ and
$\Gamma_2$ be two generic products of Dirac $\gamma$ matrices, a
two-point correlator is defined as
\begin{eqnarray}
\label{eq:correlator}
 C_{\Gamma_1,\Gamma_2}(t) &=& \sum_{\mathbf{x}}\left<
\left( \bar{u} \Gamma_1 d \right)^\dagger(t,\mathbf{x})
\left( \bar{u} \Gamma_2 d \right)(0)
\right> \comma
\end{eqnarray}
where $u$ and $d$ are the fields corresponding to two different quark flavors,
which from now on we take to be mass degenerate. All possible choices for the
non-singlet quark bilinears and the quantum numbers of the corresponding
physical states are summarized in Tab.~\ref{tab:bilinear}.
\TABLE{
  \begin{tabular}{||c|c|c|c|c|c||}
    \hline
    Particle & $a_0$ & $\pi$ & $\rho$ & $a_1$ & $b_1$ \\ \hline
		Bilinear & $\bar{u} d$ & $\bar{u} \gamma_5 d$, $\bar{u} \gamma_0 \gamma_5 d$  &
 $\bar{u} \gamma_i d$, $\bar{u} \gamma_0 \gamma_i d$ & $\bar{u} \gamma_5 \gamma_i d$ &  $\bar{u} \gamma_i \gamma_j d$  \\ \hline
		$J^{PC}$ & $0^{++}$ & $0^{-+}$ & $1^{--}$ & $1^{++}$ & $1^{+-}$ \\ 
    \hline
  \end{tabular}
  \caption{Bilinear operators for the computation of non-singlet meson masses.} 
  \label{tab:bilinear}
}

Performing the Wick contractions we can rewrite $C_{\Gamma_1,\Gamma_2}(t)$ in terms of
the quark propagator $G(x,y)=(D_m)^{-1}(x,y)$ or, equivalently, of its hermitean version
$H(x,y)=G(x,y)\gamma_5$:
\begin{eqnarray}
C_{\Gamma_1,\Gamma_2}(t) &=& - \sum_{\mathbf{x}}\left< \tr
\left[ \gamma_0 \Gamma_1^\dagger \gamma_0 G(x,0) \Gamma_2 \gamma_5 G(x,0)^\dagger \gamma_5 \right]
\right> = \nonumber \\
&=& - \sum_{\mathbf{x}}\left< \tr
\left[ \gamma_0 \Gamma_1^\dagger \gamma_0 H(x,0) \gamma_5 \Gamma_2 H(x,0) \gamma_5 \right]
\right> \comma
\end{eqnarray}
($\tr$ indicates the trace over spinor and color indices).

The propagator $G(x,0)_{AB}$ is obtained by inverting the Dirac operator $D_m$ over 
point sources (capital roman letters $A,B,\dots$ are used for collective indices over spin and color):
\begin{equation}
G(x,0)_{AB} = (D_m)^{-1}_{AC}(x,y) \delta_{CB}\delta_{y,0} =  (D_m)^{-1}_{AC}(x,y) \eta_{C}^{(B)}(y)\comma \label{propinv}
\end{equation}
where the second equality defines the $4N$ point sources $\eta^{(B)}$.

The algorithm used for the inversion in Eq.~(\ref{propinv}), is a
multishift QMR. This enables us to compute all the quark propagators
corresponding to different masses simultaneously. We use a version of
the QMR suitable for $\gamma_5$-hermitean matrices with even-odd
preconditioning of the Dirac matrix~\cite{Frommer:1995ik}.
In the rare cases when
the algorithm fails to converge, we continue the search for a solution
using the MINRES algorithm, which is guaranteed to converge, on the
hermitean version of the Dirac operator. For all the inversions we
required a relative precision of $10^{-5}$. To this accuracy, and for
the parameters given above, the number of required applications of the Dirac
operator to compute the propagator $G(x,0)_{AB}$  at fixed values for the
hopping parameters $\kappa$ is found
to become independent of $N$. We found that the average number of
applications of the Dirac matrix required is about $7500$, $5000$,
$5000$ for $N=3,4,6$ respectively (for SU(2) we used different
parameters).

From general large--$N$ arguments, we expect the occurrence of
exceptional configurations to be suppressed as $N$ increases. This is
confirmed by preliminary results reported in
Ref.~\cite{Bali:2007kt}. At the values of the parameters we have
simulated, there is no sign of the presence of exceptional
configurations.

Simulations have been performed with a bespoke code, which has been tested
against published results for SU(3) (see~\cite{Gockeler:1997fn} for a review
of the literature). We have collected 100 configurations for each value of $N$,
separated by 50 Monte Carlo sweeps.
\section{Numerical results}
\label{sec:num}
\subsection{Extracting masses from correlators}
\label{sec:num:gen}
Masses can be extracted from the large--$t$ behavior of $C_{\Gamma,\Gamma}(t)$.
Inserting the energy eigenstates in the RHS of Eq.~(\ref{eq:correlator}) yields
\begin{eqnarray}
C_{\Gamma,\Gamma}(t) = \sum_i |c_i|^2 e^{- m_i t} \ ,
\end{eqnarray}
where $c_i = (1/2 m_i) \left< 0 \left| \left(\bar{u} \Gamma d
    \right)(0) \right| i \right>$, $\Gamma$ is one of the $\gamma$
matrix products appearing in the bilinears in Tab.~\ref{tab:bilinear},
$|i \rangle$ is an eigenstate of the Hamiltonian with the same quantum
numbers as the fermion bilinear, and $m_i$ is the mass of the $|i
\rangle$ eigenstate. In the limit $t \to \infty$, the previous
equation becomes
\begin{eqnarray}
C_{\Gamma,\Gamma}(t) \mathop{=}_{t \to \infty} |c_0|^2 e^{- m_0 t} \ ,
\end{eqnarray}
i.e. at large time correlation functions decay in time as a single exponential
with a typical time
given by the inverse mass of the lowest-lying state in the spectrum
with matching quantum numbers. The lowest mass in a given channel can then be
extracted as
\begin{eqnarray}
\label{eq:mas1}
m_0 = - \lim_{t \to \infty} \frac{\log C_{\Gamma,\Gamma}(t)}{t} \ .
\end{eqnarray}
Practically, one defines the effective mass $m_0 (t)$ as
\begin{eqnarray}
\label{eq:meff1}
m_0 (t) = - \log \frac{C_{\Gamma,\Gamma}(t)}{C_{\Gamma,\Gamma}(t-1)}
\end{eqnarray}
and obtains $m_0$ by fitting $m_0 (t)$ to a constant at large enough $t$.

On a finite lattice the exponential in the large--time behavior of
the propagator is replaced by a $\cosh$ and an effective mass can be
defined as
\begin{eqnarray}
\label{eq:meff}
m_0 (t) = \mathrm{acosh}\left( \frac{C_{\Gamma,\Gamma}(t+1) + C_{\Gamma,\Gamma}(t-1)} {2 C_{\Gamma,\Gamma}(t)}\right) \ .
\end{eqnarray}
Typical examples of correlation functions and effective masses as a function
of the separation between source and sink are shown respectively in
Fig.~\ref{fig:excorrelator} and Fig.~\ref{fig:exmass}.

\FIGURE[t]{
  ~\\
  \epsfig{file=FIGS/g5_corr_su4.eps,scale=0.6}
  \label{fig:excorrelator}
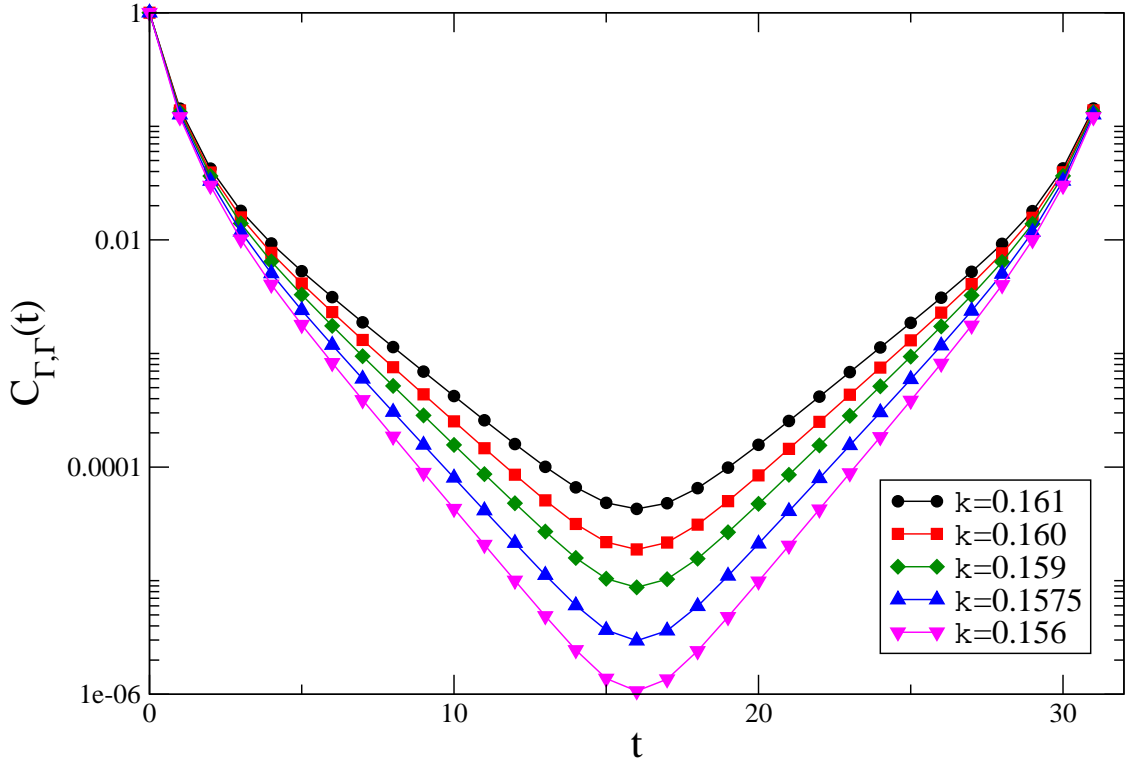
  \caption{Correlators for SU(4) and $\Gamma = \gamma_5$ at the values of
    $\kappa$ shown. The correlators have been normalized in such a way that
    at $t=0$ their value is 1. The lines joining the data are only guides
    for the eyes.}
}
\FIGURE[t]{
  \epsfig{file=FIGS/g5_meff_su4.eps,scale=0.6}
  \label{fig:exmass}
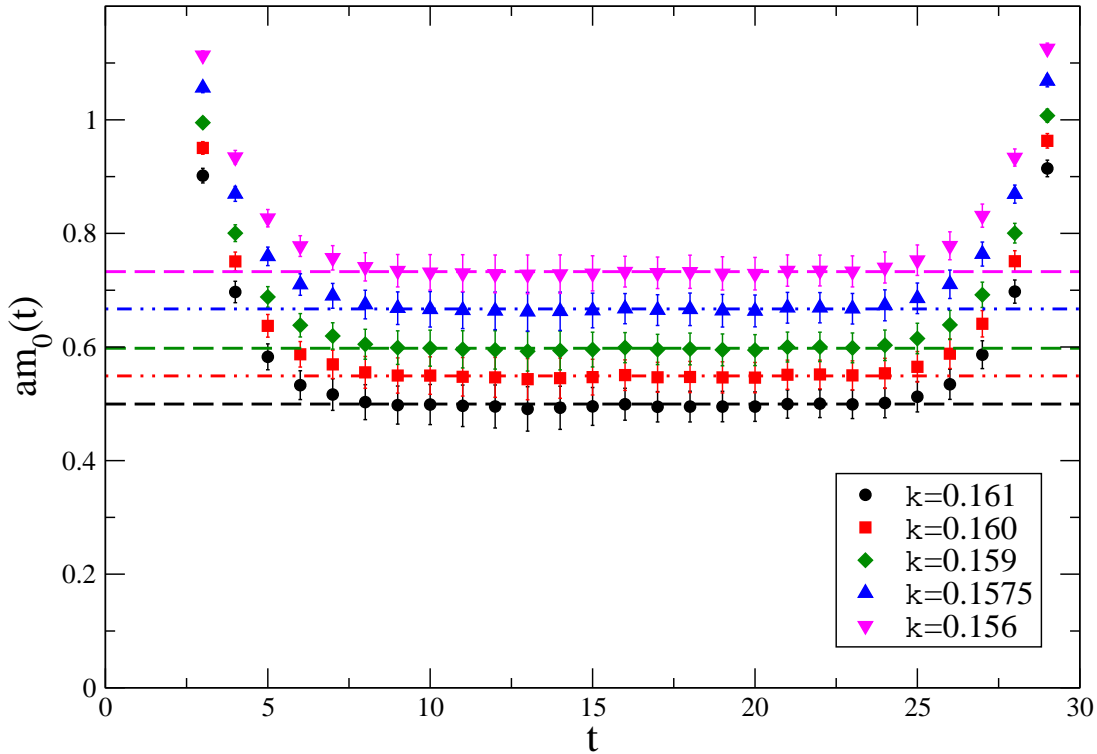
  \caption{Effective masses from the correlators in
    Fig.~\ref{fig:excorrelator}. The straight lines are fits to the data at
    plateau.}
}

We have estimated the errors on the correlators using a jack-knife
method, and checked that the bootstrap method gives similar
results. With simple link operators, we have been able to extract an
unambiguous signal for the $\pi$ and the $\rho$ mesons (to which we
limit our analysis). Other correlators yielded a noisy signal, and we
plan to investigate the possibility of improving the signal-to-noise
ratio by more sophisticated measurements.  Due to the small number of
data points, often correlated fits proved to be unreliable, as already
observed (see e.g. Refs.~\cite{Michael:1993yj,Michael:1994sz}).
Hence, masses have been extracted with uncorrelated fits and the error
estimated with a jack-knife procedure. We have checked that the
uncorrelated fit results coincide with the correlated fit results
whenever the correlated fits give reasonable values for the parameters
and the $\chi^2$.

Our results for the various $N$ for the PCAC mass $m_\mathrm{PCAC}$, the mass
of the pion $m_{\pi}$, and the mass of the $\rho$ $m_{\rho}$ are
reported in Tabs.~\ref{tab:numsu2}--\ref{tab:numsu6}. The details of
our analysis are explained in the following two subsections. 

In order to convert the results expressed in lattice units to masses in
physical units, we note that
\begin{eqnarray}
a m = (a T_c) (m/T_c) = m / (5 T_c) 
\end{eqnarray}
(the last equality uses the fact that the lattice spacing has been fixed
in such a way that the deconfinement phase transition corresponds to
$N_t = 5$). As a reference scale, we can use $T_c$ for SU(3), which is
approximately $270 \ \mathrm{MeV}$.

\begin{table}
  \begin{center}
  \begin{tabular}{|c|c|c|c|c|c|c|}
      \hline
      $\kappa$ & $a m_\mathrm{PCAC}$ & $m_\mathrm{PCAC}$ (MeV) & $a m_{\pi}$
& $m_{\pi}$ (MeV) & $a m_{\rho}$ & $m_{\rho}$ (MeV)\\
      \hline
      0.152 & 0.0824(30) & 111(4) & 0.541(3) & 730(4) & 0.620(6) & 837(8)\\
      0.153 & 0.0669(28) &  90(4) & 0.492(4) & 664(5) & 0.584(7) & 788(9)\\
      0.154 & 0.0522(25) &  70(3) & 0.441(4) & 595(5) & 0.547(7) & 738(9)\\
      0.155 & 0.0396(24) &  53(3) & 0.389(4) & 525(5) & 0.510(10) & 789(13)\\
      0.156 & 0.0261(22) &  35(3) & 0.319(7) & 430(9) & 0.458(17) & 618(23)\\
      \hline
  \end{tabular}
  \caption{Numerical results for SU(2). Masses in lattice units 
    have been converted to physical units by noting $a=1/(5T_c)$.}
  \label{tab:numsu2}
  \end{center}
\end{table}

\begin{table}
  \begin{center}
  \begin{tabular}{|c|c|c|c|c|c|c|}
      \hline
      $\kappa$ & $a m_\mathrm{PCAC}$ & $m_\mathrm{PCAC}$ (MeV) & $a m_{\pi}$
& $m_{\pi}$ (MeV) & $a m_{\rho}$ & $m_{\rho}$ (MeV)\\
      \hline
      0.156 &  0.1047(24) & 141(3) & 0.625(2) & 844(3) & 0.720(3) & 972(4)\\
      0.1575 & 0.0797(21) & 108(3) & 0.553(2) & 747(3) & 0.667(4) & 900(5)\\
      0.159 & 0.0574(17) & 77(2) & 0.476(2) & 643(3) & 0.616(5) & 832(7)\\
      0.160 & 0.0431(16) & 58(2) & 0.420(2) & 567(3) & 0.582(6) & 786(8)\\
      0.161 & 0.0299(14) & 40(2) & 0.362(3) & 489(4) & 0.550(7) & 743(9)\\
      \hline
  \end{tabular}
  \caption{Numerical results for SU(3). Masses in lattice units 
    have been converted to physical units by noting $a=1/(5T_c)$.}
  \label{tab:numsu3}
  \end{center}
\end{table}

\begin{table}
  \begin{center}
  \begin{tabular}{|c|c|c|c|c|c|c|}
      \hline
      $\kappa$ & $a m_\mathrm{PCAC}$ & $m_\mathrm{PCAC}$ (MeV) & $a m_{\pi}$
& $m_{\pi}$ (MeV) & $a m_{\rho}$ & $m_{\rho}$ (MeV)\\
      \hline
      0.156  & 0.1506(35) & 203(5) & 0.733(1) & 990(1) & 0.817(2) & 1103(3)\\
      0.1575 & 0.1234(29) & 167(4) & 0.667(1) & 900(1) & 0.766(2) & 1034(3)\\
      0.159 &  0.0981(22) & 132(3) & 0.598(1) & 807(1) & 0.714(2) & 964(3)\\
      0.160 & 0.0817(19) & 110(3) & 0.549(2) & 741(3) & 0.680(2) & 918(3)\\
      0.161 & 0.0659(17) & 89(2) & 0.499(2) & 674(3) & 0.646(3) & 872(4)\\
      \hline
  \end{tabular}
  \caption{Numerical results for SU(4). Masses in lattice units 
    have been converted to physical units by noting $a=1/(5T_c)$.}
  \label{tab:numsu4}
  \end{center}
\end{table}

\FIGURE[t]{
  \epsfig{file=FIGS/g5_meff_k0156.eps,scale=0.6}
  \label{fig:effmass_pi_all}
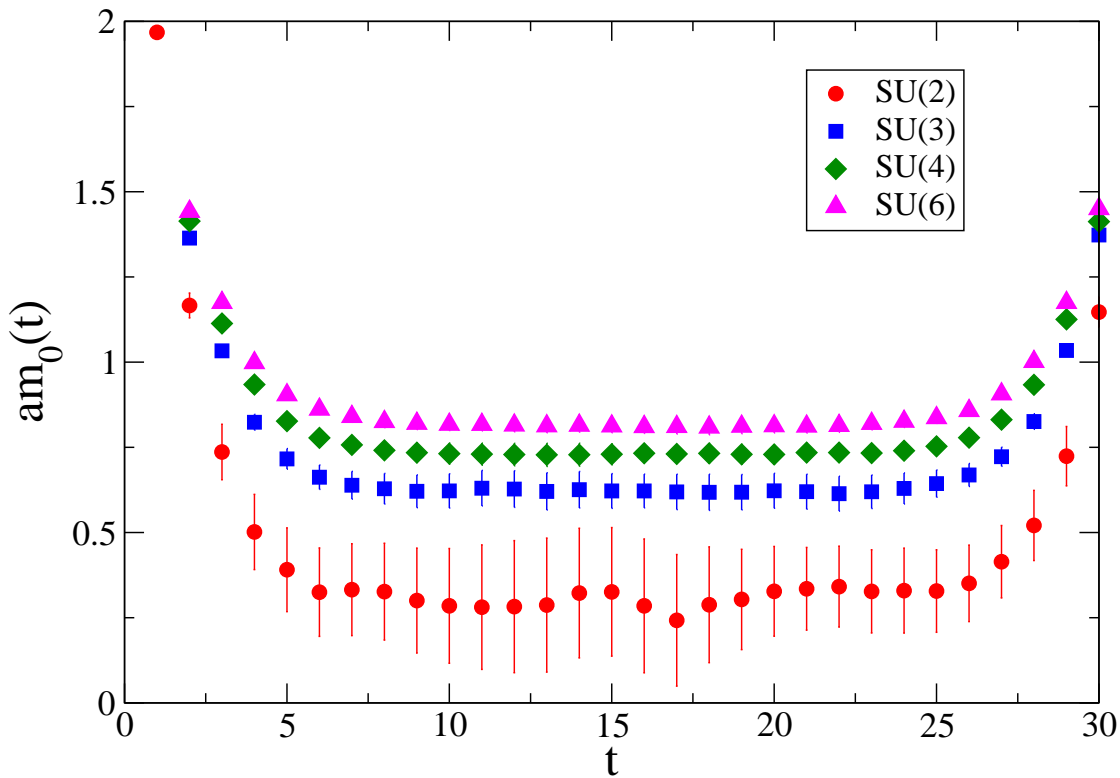
  \caption{Effective masses from $C_{\gamma_5,\gamma_5}$
    at $\kappa = 0.156$.}
}

\begin{table}
  \begin{center}
  \begin{tabular}{|c|c|c|c|c|c|c|}
      \hline
      $\kappa$ & $a m_\mathrm{PCAC}$ & $m_\mathrm{PCAC}$ (MeV) & $a m_{\pi}$
& $m_{\pi}$ (MeV) & $a m_{\rho}$ & $m_{\rho}$ (MeV)\\
      \hline
      0.156 & 0.1789(22) & 241(3) & 0.814(1) & 1099(1) & 0.889(2) & 1214(3)\\
      0.1575 & 0.1509(20) & 203(3) & 0.752(1) & 1015(1) & 0.838(2) & 1131(3)\\
      0.159 & 0.1248(18) & 168(2) & 0.687(1) & 927(1) & 0.786(2) & 1061(3)\\
      0.160 & 0.1078(17) & 146(2) & 0.6425(9) & 867(1) & 0.752(2) & 1015(3)\\
      0.161 & 0.0914(15) & 123(2) & 0.5952(9) & 804(1) & 0.718(2) & 969(3)\\
      \hline
  \end{tabular}
  \caption{Numerical results for SU(6). Masses in lattice units 
    have been converted to physical units by noting $a=1/(5T_c)$.}
  \label{tab:numsu6}
  \end{center}
\end{table}
\subsection{Meson masses at finite $N$}
\label{sec:num:pi}
The pion is the would-be Goldstone boson of chiral symmetry breaking. Chiral
perturbation theory at leading order predicts
\begin{eqnarray}
\label{eq:kappac}
m_{\pi} = A \left( \frac{1}{\kappa} - \frac{1}{\kappa_c} \right)^{1/2} \ .
\end{eqnarray} 
Hence the value of $\kappa$ corresponding to the chiral limit,
$\kappa_c$, can be obtained by fitting the pion mass according to
Eq.~(\ref{eq:kappac}). Eq.~(\ref{eq:kappac}) is modified for the
quenched theory, where quenched chiral logs appear. For quenched
SU($N$) gauge theory we expect
\begin{eqnarray}
\label{eq:chikappac}
m_{\pi} = A \left( \frac{1}{\kappa} - \frac{1}{\kappa_c} \right)^{1/[2(1+\delta)]} \ ,
\end{eqnarray} 
where $\delta$ is positive, ${\cal O}(10^{-1})$ for SU(3) and goes like 
$1/N$~\cite{Sharpe:1992ft}.

The mass of the pion can be extracted by looking at correlators
$C_{\Gamma,\Gamma}$ in which $\Gamma$ is either $\gamma_5$ or
$\gamma_0 \gamma_5$. In the latter case, it was not possible to extract
a signal for all $\kappa$'s in SU(2). Hence, although in general mass plateau
fits of $\gamma_0 \gamma_5$ have a lower $\chi^2$, we will mostly focus on
numerical results obtained with $\Gamma = \gamma_5$.

As $N$ grows, so does $m_{\pi}$ at fixed $\kappa$. A plot comparing
numerical results for the effective mass extracted from
$C_{\gamma_5,\gamma_5}$ is shown in Fig.~\ref{fig:effmass_pi_all}.  A
linear fit to the data according to Eq.~(\ref{eq:kappac}) enables us
to extract the critical values of $\kappa$ for the $N$ at which we
have simulated. Results are shown in Tab.~\ref{tab:kappac}.  The mass
obtained from the $C_{\gamma_0 \gamma_5, \gamma_0 \gamma_5}$
correlator yields compatible results. Higher statistics and a careful
study of the systematics are necessary for a more precise
determination of the critical value of $\kappa$.

Generally the reduced $\chi^2$ of the fits according to
Eq.~(\ref{eq:kappac}) varies between two and four. One can check
whether this relatively high value of $\chi^2_r$ is due to the fact
that we are neglecting chiral logarithms.  For $N \ge 2$, fits
according to~(\ref{eq:chikappac}) yield a value of $\chi^2_r$ that is
below one, but $\delta$ is found to be negative. This agrees with the
findings of~\cite{Gockeler:1997fn}, where the negative value is
interpreted as a consequence of simulating far from the chiral
limit. In fact, for $N = 2$, where we have the lightest pion mass,
$\delta$ is found to be compatible with zero. As one would have
expected, for $m_{\pi} \ge 450 \ \mathrm{MeV}$ there is no sensitivity
to the chiral logarithms~\cite{Sharpe:1992ft}.  Instead of relying on
phenomenological fits like in~\cite{Gockeler:1997fn}, we acknowledge
the impossibility to determine $\delta$ and neglect the chiral
logarithms, using the chiral behavior~(\ref{eq:chikappac}) to get an
estimate for the systematic error associated with this
approximation. For SU(2), the three-parameter fit gives a value of
$\kappa_c$ that is higher than the fit with $\delta = 0$ by about 1\%.
Considering that $\delta \propto N^{-1}$, a conservative but safe
estimate for the systematic error associated with the chiral
logarithms is of the order of a few percent.  This is in agreement
with the literature for SU(3)~\cite{Gockeler:1997fn}.  We will come
back on issues associated with the chiral logarithms in the next
subsection.

\begin{table}
  \begin{center}
    \begin{tabular}{|c|c|c|}
      \hline
      $N$ & $\kappa_c$ & $A$ \\
      \hline
      2	& 0.15827(12) &	1.0583(99)\\
      3	& 0.16359(28) &	1.142(21)\\
      4	& 0.16556(23) &	1.201(14)\\
      6	& 0.16716(12) &	1.2422(69)\\
      \hline
    \end{tabular}
    \caption{Fitted values for $\kappa_c$ and $A$ at various $N$.}
    \label{tab:kappac}
  \end{center}
\end{table}

The mass of the $\rho$ has been extracted from
$C_{\gamma_i,\gamma_i}$, after taking the average over the spatial
direction $i$ of the correlation functions.  Fits in the $C_{\gamma_0
\gamma_i,\gamma_0 \gamma_i}$ channel also yield compatible results.

At small quark mass, $m_{\rho}$ depends linearly on the quark mass and goes to
a finite value in the chiral limit. Using Eq.~(\ref{eq:kappac}), this can be
rephrased into the following relationship between $m_{\rho}$ and $m_{\pi}$
\begin{eqnarray}
\label{eq:mrhovsmpi}
m_{\rho} = m^{\chi}_{\rho} + B m_{\pi}^2 \ ,
\end{eqnarray}
where $m^{\chi}_{\rho}$ is the mass of the $\rho$ meson at the chiral point.
Note that the previous equation is not modified by chiral
logarithms~\cite{Sharpe:1992ft}. Assuming that Eq.~(\ref{eq:mrhovsmpi}) holds
in our case\footnote{More sophisticated dependencies (e.g. the addition of a
linear term in $m_{\pi}$ to Eq.~(\ref{eq:mrhovsmpi}), which is motivated by
phenomenology) are also supported by our data. In the absence of any evidence
against it, we chose to fit the parameters using the simple chiral functional
behavior.}, we can fit $m^{\chi}_{\rho}$ and $B$ at the various values of $N$
from our data. Our results for those quantities are reported in
Tab.~\ref{tab:mrhovsmpi}. The reduced $\chi^2$ of the fits (which keep into
account both the error on $m_{\rho}$ and the error on $m_{\pi}$) is always
less than one. 

\begin{table}
  \begin{center}
    \begin{tabular}{|c|c|c|}
      \hline
      $N$ & $a m^{\chi}_{\rho}$ & $B$ \\
      \hline
      2 & 0.3890(75) & 0.797(31)\\   
      3 & 0.4683(25) & 0.6455(84)\\  
      4 & 0.5018(36) & 0.5905(88)\\   
      6 & 0.5238(40) & 0.5533(77)\\  
      \hline
    \end{tabular}
    \caption{Extrapolation of $m_{\rho}$ to the chiral limit.}
    \label{tab:mrhovsmpi}
  \end{center}
\end{table}
\subsection{PCAC}
\label{sec:num:pcac}
As noted in the previous subsection, Eq.~(\ref{eq:kappac}) only holds for the
full theory, while it is modified at small masses, where quenched chiral
logs become important.  An alternative way of defining $\kappa_c$,
which is free from these ambiguities, makes use of the partially
conserved axial current (PCAC) relation. In the continuum, the PCAC
relation reads
\begin{eqnarray}
  \label{eq:pcac}
  \partial_{\mu} A^{\mu} (x) = 2 m_\mathrm{PCAC} j(x) \ ,
\end{eqnarray}
with $A^{\mu} (x) = \bar{u}(x) \gamma_{\mu} \gamma_5 d(x)$ and
$j = \bar{u}(x) \gamma_5 d(x)$. 
The previous equation allows us to determine $m_\mathrm{PCAC}$ as
\begin{eqnarray}
  m_\mathrm{PCAC} = \frac{1}{2} \frac {\langle \int \mathrm{d}{\vec{x}} (\partial_0 A^0(x)) j^{\dag}(y))  \rangle} {\langle \int \mathrm{d} {\vec{x}} j(x) j^{\dag}(y) \rangle} \ ,
\end{eqnarray}
where $y$ is an arbitrary point. On the lattice an effective mass $m_\mathrm{PCAC}(t)$ can be defined
as
\begin{eqnarray}
  \label{eq:pcacplateau}
  m_\mathrm{PCAC}(t) = \frac{1}{4} \frac{C_{\gamma_0 \gamma_5,\gamma_5} (t+1)
- C_{\gamma_0 \gamma_5,\gamma_5} (t-1) }
{C_{\gamma_5,\gamma_5} (t)} 
\end{eqnarray}
and once again fitted at plateau. Note that with our choice for the
discretized fermions PCAC holds on the lattice up to terms ${\cal O}(a)$. 
In practice, since $m_\mathrm{PCAC}(t)$ defined through Eq.~(\ref{eq:pcacplateau})
is antisymmetric around the point $N_t/2$, one averages the absolute values
at points $t$ and $N_t - t$. An example of an effective mass plateau obtained
using Eq.~(\ref{eq:pcacplateau}) is given in Fig.~\ref{fig:pcacplateau}.

\FIGURE[t]{
  \epsfig{file=FIGS/g5_g0g5_meff_k0156.eps,scale=0.6}
  \label{fig:pcacplateau}
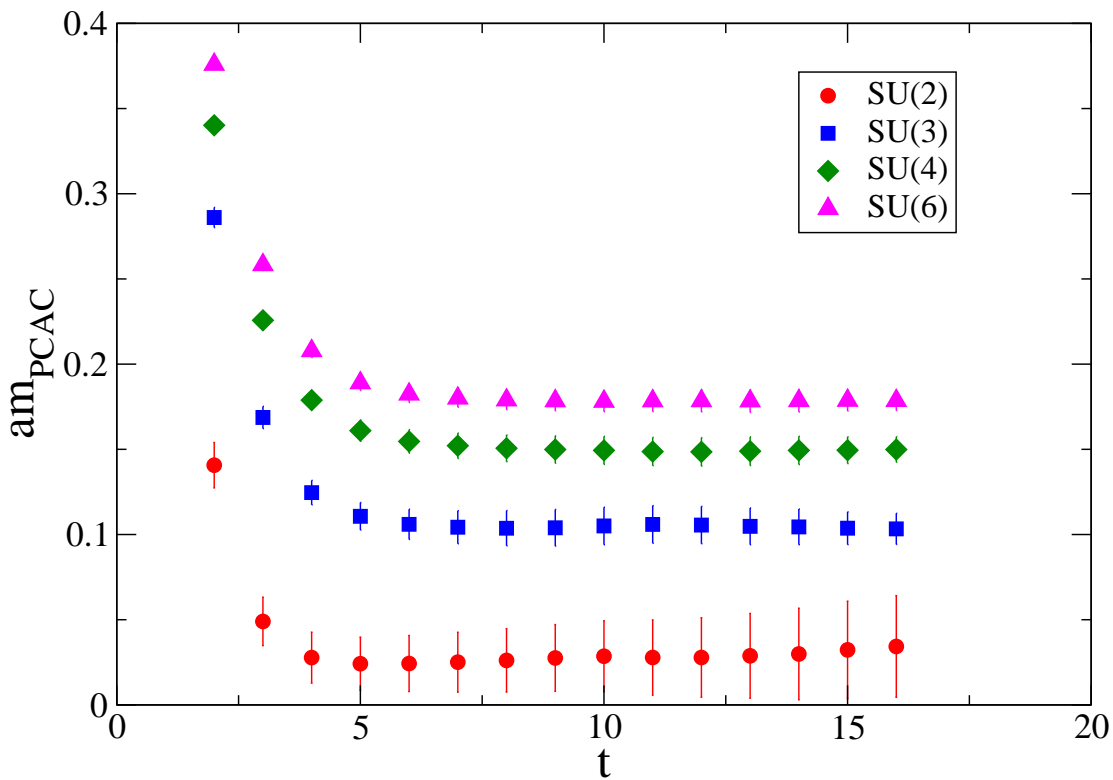
  \caption{$m_\mathrm{PCAC}(t)$ as a function of $t$ for $\kappa = 0.156$.}
}

Since $m_\mathrm{PCAC}=Z_m(1/\kappa-1/\kappa_c)$, we can determine
$\kappa_c$ as the value for which $m_\mathrm{PCAC} = 0$. A linear fit
to the data enables us to extract $\kappa_c$.  Results for $N=2,3,4,6$
are reported in Tab.~\ref{tab:kappacpcac}. Comparing with the similar
fits from $m_{\pi}$ (Tab.~\ref{tab:kappac}), it is immediate to see
that using $m_\mathrm{PCAC}$ we get values for $\kappa_c$ that are
systematically lower\footnote{This should be contrasted with fits that
keep into account chiral logarithms, for which we find values of
$\kappa_c$ systematically higher.}. Although this effect is below half
a percent, it is by far larger than the statistical errors. This
discrepancy might be due to the different chiral behavior of the two
definitions of the quark mass for the quenched theory or be a
consequence of the underestimation of the errors due to the use of
uncorrelated fits, as we have discussed above.  However, at our value
of the lattice spacing discretization errors also play a relevant
part. In order to investigate these issues, we have analyzed $m_{\pi}$
as a function of $m_\mathrm{PCAC}$. Our results are reported in
Fig.~\ref{fig:mpi_vs_mq}, where $m_{\pi}$ is plotted as a function of
$m_\mathrm{PCAC}$.

\FIGURE[t]{
  \epsfig{file=FIGS/mpi2_vs_mpcac.eps,scale=0.6}
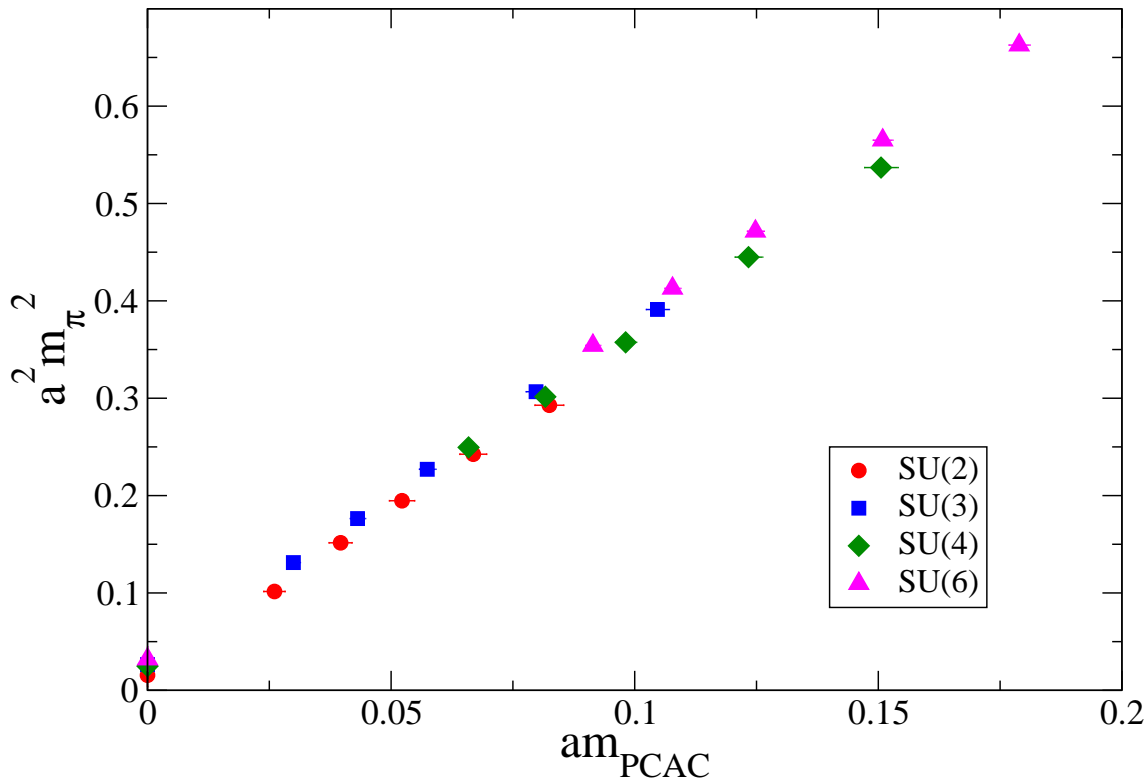
  \caption{$m_{\pi}^2$ as a function of $m_\mathrm{PCAC}$ at the various $N$.
  The values of $m_{\pi}$ at $m_\mathrm{PCAC}=0$ have been obtained with a linear
  fit to the data, as discussed in the text.}
  \label{fig:mpi_vs_mq}
}

\begin{table}
  \begin{center}
    \begin{tabular}{|c|c|c|}
      \hline
      $N$ & $\kappa_c$ & $A$\\
      \hline
      2 & 0.15792(25)  & 0.331(15)\\
      3 & 0.16306(10)  & 0.372(10)\\
      4 & 0.16513(15)  & 0.422(12)\\
      6 & 0.16657(15)  & 0.438(11)\\
      \hline
    \end{tabular}
  \end{center}
  \caption{Extrapolation of $\kappa$ to the chiral limit using $m_\mathrm{PCAC}$.}
  \label{tab:kappacpcac}
\end{table}

The expected quadratic behavior
\begin{eqnarray}
m_{\pi}^2 = C m_\mathrm{PCAC} 
\end{eqnarray}
is not obeyed by our data. To extrapolate to the chiral limit, we need
to correct the above relationship by allowing for a non-zero value for
$m_{\pi}$ when $m_\mathrm{PCAC}$ is zero:
\begin{eqnarray}
m_{\pi}^2 = C m_\mathrm{PCAC} + B \ ,
\end{eqnarray}
where $B$ and $C$ depend on $N$.  $B$ (obtained from a fit according
to the previous equation keeping into account both the errors on
$m_{\pi}$ and on $m_\mathrm{PCAC}$) is roughly of order $10^{-2}$ and
independent of $N$. If the existence of a constant (as a function of
$N$) residual $m_{\pi}$ as $m \to 0$ were a sign of the failure of the
quenched approximation, we would have expected $B$ to go to zero as $N
\to \infty$. This expectation is not supported by our data. On the
other hand, having fixed the lattice spacing across the gauge groups,
any discretization artifact would be constant in $N$. Hence, it is
likely that this residual mass is (mostly) due to the violation of
PCAC on the lattice. If this is correct, also the systematic
discrepancy between the two sets of $\kappa_c$ should be due to
lattice artifacts. In order to settle this issue, a systematic study
at different lattice spacings needs to be performed.
\section{Extrapolation to $SU(\infty)$}
\label{sec:extr}
Using data at finite $N$, we can estimate the behavior of the lowest-lying
meson masses at $N = \infty$. Following similar analysis performed in pure
gauge~\cite{Lucini:2001ej,Lucini:2003zr,Lucini:2004yh,DelDebbio:2001sj,DelDebbio:2002xa,DelDebbio:2004rw,DelDebbio:2006df}, we use predictions from the
large--$N$ expansion to see whether they hold in the non-perturbative regime.
In practice, we take the asymptotic expansion for an observable $O$ in the
quenched case~\cite{'tHooft:1973jz}
\begin{eqnarray}
\label{lnexpansion}
O(N) = O(\infty) + \sum_i \frac{\alpha_i}{N^{2i}}
\end{eqnarray}
and we check whether a reasonable (as dictated by the number of data)
truncation of this series accommodates our numerical values. In the
pure gauge case, a precocious onset of the large--$N$ behavior has
been found for all the observables that have been studied
(which include glueball masses, deconfining temperature and
topological susceptibility): the ${\cal O}(1/N^2)$ correction
correctly describes the data down to at least $N = 3$, often including
also the case $N=2$. From a qualitative point of view, it is already
clear from what we have seen so far that the quantities we have
investigated have a mild dependence on $N$.  In this section, we want
to study whether this dependence is correctly described by a
large--$N$-inspired expansion.

$\kappa_c$ can be computed in lattice perturbation theory~\cite{Stehr:1982en}.
The result at one loop is in agreement with the predictions of the large--$N$
limit: this quantity receives a correction ${\cal O}(1/N^2)$. This motivates
the fit
\begin{eqnarray}
\kappa_c(N) = \kappa_c(\infty) + \frac{a}{N^2} \ .
\end{eqnarray} 
For $\kappa_c$ obtained via Eq.~(\ref{eq:kappac}), we get
$\kappa_c(\infty) = 0.1682(1)$ and $a = -0.0398(6)$, with $\chi^2_r =
0.6$. The quality of the fit is good, and the coefficient of the
$1/N^2$ correction is small, as one would expect for a series
expansion. Similarly to the pure gauge case, we observe an early onset
of the asymptotic behavior, which captures also the SU(2) value. Our
data and the large--$N$ extrapolation are plotted in
Fig.~\ref{fig:lnkappac}. The same extrapolation for the critical value
of $\kappa$ obtained using the PCAC relation yields $\kappa_c(\infty)
= 0.1675(2)$, $a = -0.039(1)$ and $\chi^2_r = 1.3$. The discrepancy
between the values of $\kappa_c(\infty)$ could be due to lattice
discretization artifacts, as discussed in Sect.~\ref{sec:num:pcac}. We
take the difference between the two determinations should be seen as
an estimate of the systematic error.  The fact that the angular
coefficient $a$ has the same value seems to corroborate this
hypothesis. A more precise determination of $\kappa_c$ is beyond the
scope of this work. 
 
\FIGURE[t]{
  \epsfig{file=FIGS/kc_alln.eps,scale=0.6}
  \caption{Extrapolation of $\kappa_c$ to $N = \infty$.}
  \label{fig:lnkappac}
}

The slope $A$ in Eq.~(\ref{eq:kappac}) can also be extrapolated to the
$N = \infty$ limit, with corrections that are ${\cal
  O}({1/N^2})$:
\begin{eqnarray}
A(N) = A(\infty) + a/N^2 \ .
\end{eqnarray}
For the $C_{\gamma_5,\gamma_5}$ results, the fit gives $A(\infty) = 1.262(8)$
and $a = -0.82(6)$ with $\chi^2_r = 1.2$. This allows us to write the mass
of the $\pi$ as a function of $\kappa$ at $N = \infty$ as
\begin{eqnarray}
\label{eq:kappac_infty}
m_{\pi} = 1.262(8) \left( 1/\kappa - 5.945(4) \right)^{1/2} \ ,
\end{eqnarray}
where the values of $\kappa_c$ obtained from fits to $C_{\gamma_5,\gamma_5}$
have been used (using the PCAC value gives a slightly discrepant result,
for the reasons discussed in Sect.~\ref{sec:num:pcac}).
We plot in Fig.~\ref{fig:mpi_vs_kappa} the data for the dependence of
$m_{\pi}^2$ as a function of $\kappa$, with a fit according to
Eq.~(\ref{eq:kappac}). We also plot in the same figure
Eq.~(\ref{eq:kappac_infty}).

\FIGURE[t]{
  \epsfig{file=FIGS/mpi_vs_kappa.eps,scale=0.6}
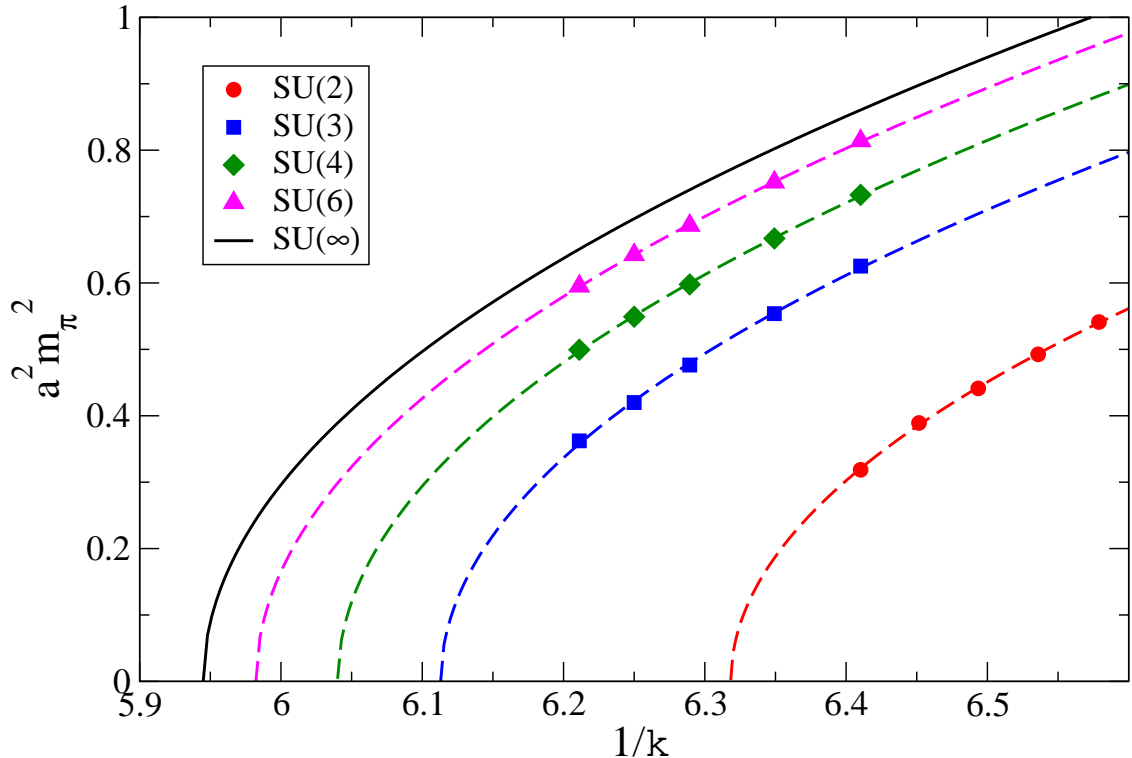
  \caption{$m_{\pi}$ as a function of $1/\kappa$. The curves through the data
  are obtained from a fit assuming the expected leading dependence from
  chiral perturbation theory. Also shown is the extrapolation to $N = \infty$.}
  \label{fig:mpi_vs_kappa}
}

The parameters describing $m_{\rho}$ as a function of $m_{\pi}$
(Eq.~(\ref{eq:mrhovsmpi})) are also expected to follow the asymptotic
expansion~(\ref{lnexpansion}). A fit with only the leading $1/N^2$ correction
yields
\begin{eqnarray}
\begin{array}{ll}
m^{\chi}_{\rho}(N) = 0.539(3)  - 0.62(3)/N^2 \ ,& \qquad \chi^2_r = 0.008 \ ;\\
B(N) = 0.5224(8) + 1.10(1)/N^2  \ , & \qquad \chi^2_r = 0.7 \ .
\end{array}
\end{eqnarray}
The tiny $\chi^2_r$ for $B(\infty)$ is particularly surprising,
given the statistical independence of the measured values of
$B(N)$. We see that once again the leading behavior describes very well the
parameters and that the coefficient of the $1/N^2$ correction is order one. 

As a result of this analysis, at $N = \infty$ we can describe $m_{\rho}$ as
a function of $m_{\pi}$ as 
\begin{eqnarray}
m_{\rho} = 0.539(3) + 0.5224(8) m_{\pi}^2 \ .
\end{eqnarray}
This relationship, together with the data and the fits at finite $N$, is
plotted in Fig.~\ref{fig:mpi_vs_mrho2}.

\FIGURE[t]{
  \epsfig{file=FIGS/mpi_vs_mrho2.eps,scale=0.6}
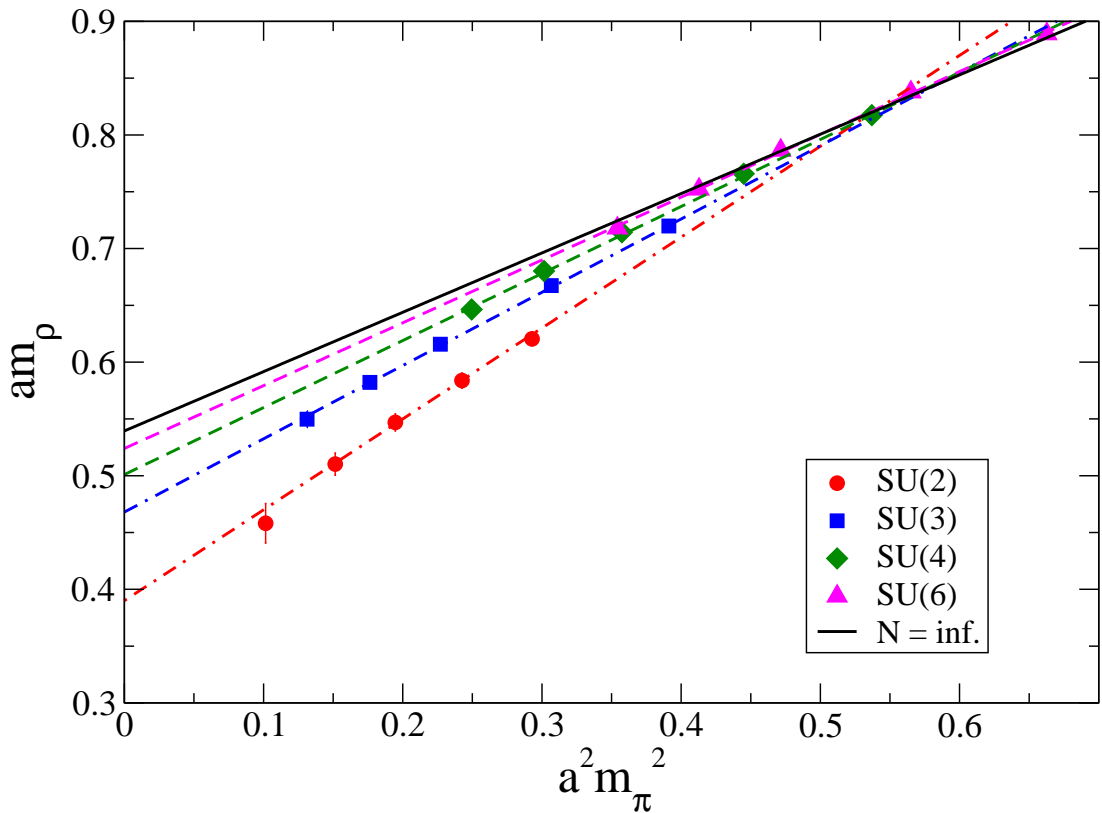
  \caption{$m_{\rho}$ vs. $m_{\pi}$. Lines through the data have
  been obtained with a plot inspired by chiral perturbation theory. Also shown
  is the extrapolation to $N = \infty$.}
  \label{fig:mpi_vs_mrho2}
}

\section{Discussion and conclusions}
\label{sec:dis}
In this paper, we have used standard lattice QCD methods for computing
correlation functions to extract the masses of the $\rho$ and $\pi$
mesons in the large--$N$ limit of SU($N$) gauge theories. The masses in
the limiting theory have been obtained with an extrapolation from the
quenched data at $N = 2, 3, 4, 6$ using the large--$N$ behavior deduced from
arguments inspired by a diagrammatic expansion. We find that the extrapolation
works well in its simplest form, i.e. using only the leading correction
to the large--$N$ value. This allows us to determine the
behavior at $N = \infty$ of the mass of the pion as a function of the
renormalized quark mass and of the mass of the $\rho$ as a function of the
mass of the pion (chiral perturbation theory has also been used as an input).
The two central results of this paper are summarized by the parametrization
\begin{eqnarray}
\label{largeN1}
m_{\pi}(N) = \left(1.262(8) - \frac{0.82(6)}{N^2}\right)
\left( \frac{1}{\kappa} - 5.945(4) + \frac{0.0398(6)}{N^2} \right)^{1/2}
\end{eqnarray}
and 
\begin{eqnarray}
\label{largeN2}
m_{\rho}(N) =  \left(0.539(3)  - \frac{0.62(3)}{N^2}\right) +
\left(0.5224(8) + \frac{1.10(1)}{N^2}\right) m_{\pi}^2(N) \ .
\end{eqnarray}

One of the motivations to perform a calculation from first principles for
the large--$N$ limit of SU($N$) gauge theories is to compare with predictions
obtained in the AdS/CFT framework. To date, no AdS background has been found
that can be considered a good dual description of non-supersymmetric QCD
extrapolated at large--$N$. Hence, one could use our calculation to benchmark
the proposed AdS ansatz. From this point of view, we notice a striking
agreement with a calculation~\cite{Babington:2003vm,Erdmenger:2007cm} using
the Constable-Myers background~\cite{Constable:1999ch},
which (after normalizing the mass of the
$\rho$ in the chiral limit to our data) finds for the coefficient of
$m_{\pi}^2$ (see Eq.~(\ref{eq:mrhovsmpi})) $0.57$.
This number is in agreement with our calculations within 5\%. However,
before we can draw any conclusion from the comparison with the lattice, an
extrapolation of the lattice data to the continuum limit is needed.

Beyond the specific numerical details, our calculation seems to
indicate that (a) the large--$N$ theory is a well defined theory; (b)
lattice calculations can be successfully exploited to compute the
parameters of this theory; (c) at least in the quenched theory, to
describe results at any finite $N$ only the first term in the expected
power series in $1/N^2$ is required; (d) the coefficient of the
correction is at most order one, justifying the idea of a power
expansion.  All these indications are perfectly in line with what we
have already learned for the SU($N$) theory without fermionic
matter. Although this can be considered obvious, since our calculation
is quenched, we stress that the $N = \infty$ limit is also
quenched. In other words, to describe the limiting theory a quenched
calculation suffices. The inclusion of the full fermion determinant
becomes mandatory if one is interested in the actual size of the
finite $N$ corrections. In particular, one expects larger corrections
(${\cal O}(n_f/N)$) in the unquenched theory.

As we have stressed several times, one of the main limitations of this
calculation is that our chiral extrapolations are not sensitive to the
expected chiral log behavior. We have conservatively estimated that
this approximation produces a 3\% systematic error. This error does not
affect our conclusions. Moreover, we note that chiral logarithms do not
modify Eq.~(\ref{largeN2}). In any case, in order to make more robust
estimates, better control on the chiral extrapolation should be 
achieved. This requires simulating at smaller pion masses. It would
also be nice to check that the chiral log effects decrease as $N$
increases. 

Another source of systematic error in our calculation is the fact that
simulations have been performed at one single lattice spacing. For this
reason we regard
our results as exploratory. As discussed in Sect.~\ref{sec:num:pcac},
lattice artifacts do seem to play a role, although they are not big enough
to spoil the features of the theory and the way in which the large--$N$
behavior is approached. Nevertheless, a study closer to the continuum
is necessary to clarify these issues. Work for the extrapolation to the
continuum limit is already in progress. For this extrapolation, the use of
an improved fermion action can mitigate the discretization artifacts,
decreasing considerably the required numerical effort. We shall explore this
possibility in the future. As for finite size effects, we have argued in
Sect.~\ref{sec:lat} that with our choice of parameters they can be neglected,
but this also ought to be verified directly.

Aside from technicalities, other features of the large--$N$ theory, like
the spectrum in the flavor singlet channel and masses of heavier mesons, also
deserve to be investigated. While the latter problem can probably be
dealt with using improved techniques for computing correlation functions
(e.g.~with smeared links replacing straight links and smeared sources replacing
point sources), for the scalar mesons,
for which disconnected contributions are important, different strategies
need to be adopted. An adaptation of the techniques exposed
in~\cite{Foley:2005ac,Collins:2007mh} is in progress. Results will be reported
in a future publication.

\section*{Acknowledgments}
We thank C. Allton, G. Bali, F. Bursa and C.~Nu\~nez for useful comments
on this manuscript.
Discussions with A. Armoni, N. Dorey, N. Evans, L. Lellouch, P.
Orland, G. Semenoff, M. Shifman and G. Veneziano are gratefully acknowledged.
Numerical simulations have been performed on a 60 core Beowulf cluster
partially funded by the Royal Society and STFC. We thank SUPA for financial
support for the workshop {\em Strongly interacting dynamics beyond
the Standard Model}, during which many of the ideas reported in this paper
were discussed. B.L. and L.D.D. thank the
hospitality of the Isaac Newton Institute, where this work was finalized.
L.D.D. is supported by an STFC Advanced Fellowship. B.L. is supported by
a Royal Society University Research Fellowship. The work of C.P. has been
supported by contract DE-AC02-98CH10886 with the U.S. Department of Energy.

\bibliography{largeN,hirep,alg,mesons}
\end{document}